# Spectral Analysis of Dow Jones Index and Comparison with Model Predicted Cycles During 1900-2005

A. M. Selvam

Deputy Director (Retired)

Indian Institute of Tropical Meteorology, Pune 411 008, India

Email: amselvam@gmail.com

Web sites: http://www.geocities.com/amselvam

http://amselvam.tripod.com/index.html

## Abstract

The day-to day fluctuations of Dow Jones Index exhibit fractal fluctuations, namely, a zigzag pattern of successive increases followed by decreases on all space-time scales. Self-similar fractal fluctuations are generic to dynamical systems in nature and imply long-range space-time correlations. The apparently unpredictable (chaotic) fluctuations of dynamical systems exhibit underlying order with the power spectra exhibiting inverse power law form, now identified as self-organized criticality. The physics of self-organized criticality is not yet identified. A general systems theory developed by the author shows that self-similar fractal fluctuations are signatures of quantum-like chaos in dynamical systems of all size scales ranging from the subatomic dynamics of quantum systems to macro-scale fluid flows. The model predicts the universal inverse power-law form of the statistical normal distribution for the power spectra of fractal space-time fluctuations of dynamical systems. In this paper it is shown that the power spectrum of 100 years of normalized month to month fluctuations of Dow Jones index exhibits the universal inverse power law form of the statistical normal distribution consistent with model prediction. It is shown that prediction of times of occurrence of maxima and minima during the two years subsequent to the data period used for the study is possible using the dominant peak periodicities obtained from the continuous periodogram spectral analysis of historic data.

## 1. Introduction

Dow Jones Index, an indicator of aggregate stock market activity exhibits irregular fractal fluctuations and exact prediction is difficult. Such fractal fluctuations generic to dynamical systems in nature such as fluid flows, heartbeat patterns, rainfall, population growth, etc., is now (since 1980s) a multidisciplinary intensive field of study, namely, *complexity and chaos* (Gleick, 1987). Fractal fluctuations are self-similar characterized by successive increases followed by decreases on all space-time scales. The power spectra of fractal fluctuations follow inverse power law of form $f^{\alpha}$ where $f$ is the frequency and the scale factor $\alpha$ varies for different frequency ranges, implying multifractal fluctuation structure. Self-similarity implies long-range space-time correlations and is identified as self-organized criticality (Bak, Tang, Wiesenfeld, 1987, 1988; Bak and Chen, 1991, Turcotte and Rundle, 2002). The physics of self-organized criticality is not yet identified. Selforganized criticality represents an attempt to extend the universality and scaling of the critical point to many-body systems far from equilibrium (McCauley, 2004).

Stanley *et al* (2002) discus some of the similarities between work being done by economists and by physicists seeking to contribute to economics. They also mention some of the differences in the approaches taken and seek to justify these different approaches by developing the argument that by approaching the same problem from different points of view, new results might emerge. In particular, they review two newly discovered scaling results that appear to be universal, in the sense that they hold for widely different economies as well as for different time periods: (i) the fluctuation of price changes of any stock market is characterized by a probability density function, which is a simple power law with exponent -4 extending over $10^2$ SDs (a factor of $10^8$ on the y axis); this result is analogous to the Gutenberg-Richter power law describing the histogram of earthquakes of a given strength; and (ii) for a wide range of economic organizations, the histogram shows how size of organization is inversely correlated to fluctuations in size with an exponent $\approx 0.2$. Neither of these two new empirical laws have a firm theoretical foundation.

Power laws appear to describe histograms of relevant financial fluctuations, such as fluctuations in stock price, trading volume and the number of trades. Surprisingly, the exponents that characterize these power laws are similar for different types and sizes of markets, for different market trends and even for different countries - suggesting that a generic theoretical basis may underlie these phenomena (Gabaix *et al*., 2003).

An empirical investigation about the possibility that the market is in a self-organized critical state (SOC) show a power law behaviour in the avalanche size, duration and laminar times during high activity period (Bartolozzi, Leinweber and Thomas, .2005).

**1.1 Stock market fractal fluctuations, Elliot waves and Fibonacci series**

The self-similar fluctuations in stock market price fluctuations was identified in the 1930s as waves within waves by R. N. Elliot (Elliot, 1938; Frost and Prechter, 1998) who also found that the successive waves are in relation with each other through the Fibonacci number series. The golden mean underlies Fibonacci number series. The golden mean is the most irrational number and is associated with the Fibonacci mathematical sequence 1, 1, 2, 3, 5, 8, ...... where each term is the sum of the two previous terms and the ratio of each term to the previous term approaches the golden mean $\tau$ equal to $(1+\sqrt{5})/2 \approx 1.618$.

That the apparently irregular fractal patterns are governed by the precise geometry of the Fibonacci number series was identified more than 150 years ago in Botany (Arber, 1950; Jean, 1994). The branching (bifurcating) structure of roots, shoots, veins on leaves of plants, etc., have similarity in form to branched lighting strokes, tributaries of rivers, physiological networks of blood vessels, nerves, etc. (Goldberger *et al*., 1990, 2002; Jean, 1994). Such seemingly complex network structure is associated with exquisitely ordered beautiful patterns exhibited in flowers and arrangement of leaves in the plant kingdom (Jean, 1994). The botanical elements which constitute plants, such as branches, leaves, petals, florets, etc, begin their existence as primordia in the neighborhood of the undifferentiated shoot apex (extremity). Extensive observations in botany show that in more than 90% of plants studied worldwide (Jean, 1994; Stewart, 1995) primordia emerge as protuberances at locations such that the angle subtended at the apical center by two successive primordia is equal to the golden angle $\phi=2\pi/\tau^2$ corresponding to approximately 137.5° where $\tau$ is the golden mean $(1+\sqrt{5})/2 \approx 1.618$. The surprisingly precise geometrical placement of plant primordia results in the observed 'phyllotactic patterns', namely, the familiar spiral patterns found in the arrangement of leaves on a stem, in florets of composite flowers, the pattern of scales on pineapple and pine cone, etc.

A general systems theory for fractal fluctuations (Selvam, 1990, 1998; Selvam and Fadnavis 1998) summarized in Section 2 shows mathematically that the golden mean underlies fractal fluctuations.

## 1.2 Mathematical models for fractal fluctuations and deterministic chaos

The larger scale fluctuations incorporate the smaller scale fluctuations as internal fine scale structure and may be visualized as a continuum of eddies (waves). Dynamical systems in nature are basically fuzzy logic systems integrating a unified whole communicating network of input signals (perturbations) with self-organized ordered two-way information transport between the larger and smaller scales. The resulting irregular fractal fluctuations are unpredictable. It has not been possible to formulate closed set of governing equations, i. e., simple mathematical equations with algebraic solutions to model and predict the fractal fluctuations of dynamical systems. Self-similar structures are generated by iteration (repetition) of simple rules for growth processes on all scales of space and time. Such iterative processes are simulated mathematically by numerical computations such as $X_{n+1} = F(X_n)$ where $X_{n+1}$, the value of the variable $X$ at $(n+1)^{th}$ computational step is a function F of its earlier value $X_n$. Mathematical models of real world dynamical systems are basically such iterative computational schemes implemented on finite precision digital computers. Computer precision imposes a limit (finite precision) on the accuracy (number of decimals) for numerical representation of $X$. Since $X$ is a real number (infinite number of decimals) finite precision introduces round-off error in iterative computations from the first stage of computation. The model iterative dynamical system therefore incorporates round-off error growth (Selvam, 1993). The computed solutions are sensitively dependent on initial conditions, a signature of deterministic chaos, so named since deterministic mathematical equations result in chaotic solutions. The computed chaotic growth patterns also exhibit self-similar fractal structure. Therefore, realistic simulation and prediction of space-time fluctuations of dynamical systems requires alternative concepts for physical laws and formulation of governing equations with algebraic solutions.

A recently proposed general systems theory attributes quantum-like chaos to the observed long-range correlations of fractal fluctuations. The model predicts: (a) the universal inverse power law form of the statistical normal distribution for the power spectra of fractal fluctuations (b) observed fractal fluctuations result from the superimposition of component eddies (c) identification of dominant periodicities in long-term historical data may help prediction of major cyclical changes in fluctuations in the near future. In the present study 100 years normalized monthly mean Dow Jones index was subjected to continuous periodogram power spectral analyses. The power spectrum follows universal inverse power law form of the statistical normal distribution consistent with model prediction. There is good agreement between observed and predicted times of occurrences of maxima and minima estimated from dominant periodicities identified from the 100 years historical data set.

In the following, Section 2 gives a brief summary of the General Systems Theory with applications to prediction; Section 3 presents details of Dow Jones Index data set used for the study and analyses techniques; Section 4 contains results and discussion and Section 5 gives conclusions.

## 2. Model concepts

A general systems theory was first developed to quantify the observed fractal space-time fluctuations in turbulent fluid flows. The model is based on Townsend's concept (Townsend, 1956) that large eddy structures form in turbulent flows as envelopes of enclosed turbulent eddies. The model concepts are independent of the exact details of the physical, chemical or other properties and therefore applicable to all dynamical systems. The observed inverse power law form for power spectra implies that the fractal fluctuations can be visualized to result from a hierarchical eddy continuum structure for the overall pattern. Starting from this simple basic concept that large scale eddies form as envelopes of internal small scale eddy circulations the following important model predictions are derived.

### 2.1 Quantumlike chaos in dynamical systems

Since the large scale is but the integrated mean of enclosed smaller scales, eddy energy spectrum follows statistical normal distribution according to Central Limit Theorem (Ruhla, 1992). Such a result that the additive amplitudes of eddies, when squared (variance), represents the probability densities, is observed in the subatomic dynamics of quantum systems such as the electron or photon (Rae and Harring, 1994, Maddox, 1993; Chown, 2004). The power spectra of fractal fluctuations therefore exhibit quantum-like chaos. The unified communicating network of fluctuations on all scales represents a quantum system which responds as a unified whole to local perturbation, also termed non-local connection or action at a distance. The quantum system exhibits long-term memory of short-term fluctuations.

### 2.2 Dynamic memory (information) circulation network

The root mean square (r. m. s.) circulation speeds $W$ and $w_*$ of large and turbulent eddies of respective radii $R$ and $r$ are related as

$$W^2 = \frac{2}{\pi}\frac{r}{R}w_*^2 \tag{1}$$

Equation (1) is a statement of the law of conservation of energy for eddy growth and implies a two-way ordered energy flow between the larger and smaller scales. Microscopic scale perturbations are carried permanently as internal circulations of progressively larger eddies. Fractal fluctuations therefore act as dynamic memory circulation networks with intrinsic long-term memory of short-term fluctuations.

Many complex systems in nature can be described in terms of networks capturing the intricate web of connections among the units they are made of. The key question is how to interpret the global organization of such networks as the structural subunits, (such as functionally related proteins, industrial sectors and groups of people) associated with more highly interconnected parts ( Palla *et al*., 2005).

## 2.3 Quasicrystalline structure

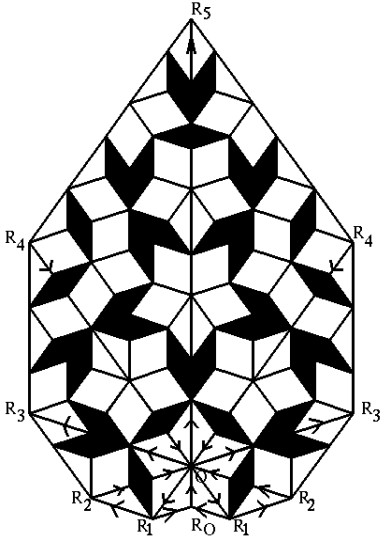

Figure 1: The quasiperiodic Penrose tiling pattern which forms the internal structure of large eddy circulations

The flow structure consists of an overall logarithmic spiral trajectory with Fibonacci winding number and quasiperiodic Penrose tiling pattern for the internal structure (figure 1). Primary perturbation $OR_O$ of time period $T$ generates return circulation $OR_1R_O$ which, in turn, generates successively larger circulations $OR_1R_2$, $OR_2R_3$, $OR_3R_4$, $OR_4R_5$, etc., such that the successive radii form the Fibonacci mathematical number series, i. e., $OR_1/OR_O = OR_2/OR_1 = \ldots = \tau$ where $\tau$ is the golden mean equal to $(1+\sqrt{5})/2 \approx 1.618$. The flow structure therefore consists of a nested continuum of vortices, i. e., vortices within vortices.

## 2.4 Dominant periodicities

Dominant quasi-periodicities $P_n$ corresponding to the internal circulations (figure 1) $OR_OR_1$, $OR_1R_2$, $OR_2R_3$, ..... are given as

$$P_n = T_s(2+\tau)\tau^n \qquad (2)$$

The dominant quasi-periodicities are equal to 2.2 $T_s$, 3.6 $T_s$, 5.8 $T_s$, 9.5 $T_s$, 15.3 $T_s$, 24.8 $T_s$, 40.1 $T_s$, 64.9 $T_s$, 105.0 $T_s$, 167.0 $T_s$, 275 $T_s$, 445.0 $T_s$ ......for values of $n$ = -1, 0, 1, 2, up to 10 respectively (Equation 2). Space-time integration of turbulent fluctuations results in robust broadband dominant periodicities (wavelengths), which are functions of the primary perturbation time period $T_s$ alone and are independent of exact details (chemical, electrical, physical etc.) of turbulent (chaotic) fluctuations. Also, such global scale oscillations in the unified network are not affected appreciably by failure of localized micro-scale circulation networks.

Wavelengths (or periodicities) close to the model predicted values have been reported in weather and climate variability (Selvam and Fadnavis, 1998), daily incidence of acute myocardial infarction (Selvam *et al.*, 2000), prime number distribution (Selvam, 2001a), Riemann zeta zeros (non-trivial) distribution (Selvam, 2001b), Drosophila DNA base sequence (Selvam, 2002), stock market economics (Selvam, 2003), Human chromosome 1 DNA base sequence (Selvam, 2004).

Similar unified communication networks may be involved in biological and physiological systems such as the brain and heart, which continue to perform overall functions satisfactorily in spite of localized physical damage. Structurally stable network configurations increase insensitivity to parameter changes, noise and minor mutations (Kitano, 2002).

## 2.5 Universal spectrum of fluctuations

Conventional power spectral analysis will resolve such a logarithmic spiral flow trajectory as a continuum of eddies (broadband spectrum) with a progressive increase in phase angle.

The power spectrum, plotted on log-log scale as variance versus frequency (period) will represent the probability density corresponding to normalized standard deviation $t$ equal to $(\log L/\log T_{50}) -1$ where $L$ is the period in units of time (month in the present study) and $T_{50}$ is the period up to which the cumulative percentage contribution to total variance is equal to 50. The above expression for normalized standard deviation $t$ follows from model prediction of logarithmic spiral flow structure and model concept of successive growth structures by space-time averaging. Fluctuations of all scales therefore self-organize to form the universal inverse power law form of the statistical normal distribution.

The normalized standard deviation values $t$ corresponding to cumulative percentage probability density $P$ equal to 50 is equal to 0 from statistical normal distribution characteristics. Since $t$ represents the eddy growth step $n$ (Equation 2) the dominant period (or length scale) $T_{50}$ up to which the cumulative percentage contribution to total variance is equal to 50 is obtained from Equation 2 for corresponding value of $n$ equal to 0. In the present study of *fractal* fluctuations of Dow Jones Index, the primary perturbation length scale $T_s$ is equal to *unit time interval* (one month in the present study) and $T_{50}$ is obtained as

$$T_{50} = (2+\tau)\tau^0 \cong 3.62 \text{ unit time interval} \qquad (3)$$

## 3. Applications of the general systems theory concepts to Dow Jones Index time series

The following model predictions for dynamical systems in general are applicable to the apparently irregular (unpredictable) fluctuations of Dow Jones Index time series and may help quantify the overall pattern and also predict times of major maxima and minima: (i) power spectra of fractal fluctuations generic to dynamical systems exhibit the universal inverse power law form of the statistical normal distribution (ii) fractal fluctuations may be resolved into an overall logarithmic spiral trajectory with the quasiperiodic Penrose tiling pattern for the internal structure (figure 1) (iii) continuous periodogram power spectral analysis therefore exhibits an eddy continuum with embedded dominant wavebands whose peak periodicities are functions of the golden mean and the primary perturbation time period (Equation 2). Peak amplitudes and phases of the statistically significant dominant eddies obtained from the periodogram analysis may help estimate the times of occurrence of maxima and minima in the near future.

## 4. Data and Analysis

Dow Jones Index values were obtained from Dow Jones Industrial Average History File, Dow Jones closing prices starting in 1900: 3 Jan 1900 to 9 November 2005 (28889 trading days). Data from: Department of Statistics at Carnegie Mellon Univ., (http://www.stat.cmu.edu/cmu-stats) at the data link http://www.analyzeindices.com/dowhistory/djia-100.txt .

The normalized day- to- day changes in the Dow Jones Index values were computed as percentages of the earlier day value. Monthly mean values were then computed from the normalized day- to- day changes in the Dow Jones Index. A total of 1266 monthly mean values were used for the study. Continuous periodogram power spectral analysis of the first 1200 (January 1900 - April 2000) values shows that the power spectrum follows the universal inverse power law form of the statistical normal distribution. Dominant peak periodicities identified from the periodogram analysis were used to estimate the Dow Jones Index values for months 1201 (May 2000) to 1266 (Oct 2005).

**4.1 Continuous periodogram power spectral analyses**

The broadband power spectrum of space-time fluctuations of dynamical systems can be computed accurately by an elementary, but very powerful method of analysis developed by Jenkinson (1977) which provides a quasi-continuous form of the classical periodogram allowing systematic allocation of the total variance and degrees of freedom of the data series to logarithmically spaced elements of the frequency range (0.5, 0). The periodogram is constructed for a fixed set of 10000 ($m$) periodicities $L_m$ which increase geometrically as $L_m=2 \exp(Cm)$ where $C=.001$ and $m=0, 1, 2,....9999$. The data series $Y_s$ for the N data points was used. The periodogram estimates the set of $A_m \cos(2\pi f_m S - \phi_m)$ where $A_m$, $f_m$ and $\phi_m$ denote respectively the amplitude, frequency and phase angle for the $m^{th}$ periodicity and $S$ is the time interval in days, months or years. Since the frequency band 0 to 0.5 accounts for N degrees of freedom, then $\nu_m$, the degree of freedom for the $m^{th}$ wavelength, is given by

$$\nu_m = 2NCf_m = \frac{2NC}{L_m}$$

except for $m=0$, when $\nu_0 = \frac{1}{2}NC$. Thus the set of $\nu_m$ is

$$\frac{1}{2}NC, NC\exp(-C), NC\exp(-2C) \text{ etc}$$

These are fixed proportions of N for a given tuning C and the sum to infinity is

$$\frac{NC}{1-\exp(-C)} - \frac{1}{2}NC = N$$

The $\chi^2$ contribution for each wavelength is given by

$$\chi_m^2 = \nu_m H_m$$

where $H_m$ is the normalized variance. The test of significance for the waveband $m = j$, $j+1, ....k$ can be made as a $\chi^2$ test for $\sum_{j}^{k} \chi_m^2$ with $\sum_{j}^{k} \nu_m$ degrees of freedom.

The cumulative percentage contribution to total variance was computed starting from the high frequency side of the spectrum. The period $T_{50}$ at which 50% contribution to total variance occurs is taken as reference and the normalized standard deviation $t_m$ values are computed as

$$t_m = \frac{\log L_m}{\log T_{50}} - 1 \qquad (4)$$

8The cumulative percentage contribution to total variance and the corresponding $t$ values were computed. The power spectra were plotted as cumulative percentage contribution to total variance versus the normalized standard deviation $t$ as given above. The period $L$ is in time interval units (months). The statistical chi-square test (Spiegel, 1961) was applied to determine the 'goodness of fit' of variance spectra with statistical normal distribution.

## 4.2 Spectral projection of near future values of Dow Jones Index

The spectral estimates are

$$Y_s = A_m \cos(2\pi f_m S - \phi_m) \tag{5}$$

where the time $S = 1$ at time $Q_1$, the first month of the series, and $L_m = 1/f_m$ goes from 2 to $m$. Just as the Fourier harmonics, $\frac{1}{2}N$ of them, each with 2 degrees of freedom (ν), at $f_m = \frac{1}{N}, \frac{2}{N}, \frac{3}{N}, \ldots, 0.5$ sum to give the actual values of $Y_s$ at time $S = 1, 2, 3,$ …..N, so closely does the summation, $\frac{1}{2}\sum_{m=0}^{\infty} \nu_m Y_{sm}$, and indeed the summation over the waveband (0.5, .001) from the 10,000 terms

$$\frac{1}{2}\sum_{0}^{9999} \nu_m A_m \cos(2\pi f_m \times 1 - \phi_m) \approx Y_1$$

$$\frac{1}{2}\sum_{0}^{9999} \nu_m A_m \cos(2\pi f_m \times 2 - \phi_m) \approx Y_2$$

$$\frac{1}{2}\sum_{0}^{9999} \nu_m A_m \cos(2\pi f_m \times 3 - \phi_m) \approx Y_3$$

$$\ldots\ldots\ldots\ldots\ldots$$

$$\frac{1}{2}\sum_{0}^{9999} \nu_m A_m \cos(2\pi f_m \times N - \phi_m) \approx Y_N$$

If the first date (month) is $Q_1$, then the date $Q$ is at time $S = (1+ Q- Q_1)$, and at auto projection from the spectral summation will give $Y$ at date (month) $Q$,

$$\frac{1}{2}\sum_{0}^{9999} \nu_m A_m \cos\left\{\left(\frac{2\pi}{L_m}(1+Q-Q_1) - \phi_m\right)\right\} \tag{6}$$

This will always give the mean value, or zero, since differences from the mean have been used. But if summation is done for values of normalized variance $H \geq H_c$, the significant part of the spectrum should remain. If $H_c = 1$, more than 70% of the variance is accounted for, and by summing $\frac{1}{2}\nu_m Y_{sm}$ for $H \geq 1$, a good fit to the data months is obtained while retaining the major significant waves in the forecast months.

## 5. Results and Discussion

The first 1200 out of the available 1266 values of monthly mean normalized Dow Jones Index (Section 4) was subjected to continuous periodogram power spectral analysis (Jenkinson, 1977). The power spectrum, plotted as cumulative percentage contribution to total variance versus normalized standard deviation $t$ is shown at figure 2 along with the statistical normal distribution and $T_{50}$ value. The power spectrum follows the model predicted universal inverse power law form of the statistical normal distribution, the 'goodness of fit' being significant (statistical $\chi^2$ test) at less than 5% level. The value of $T_{50}$, the period up to which the cumulative percentage contribution to total variance is equal to 50, is 4.38 months close to model predicted value of 3.62.

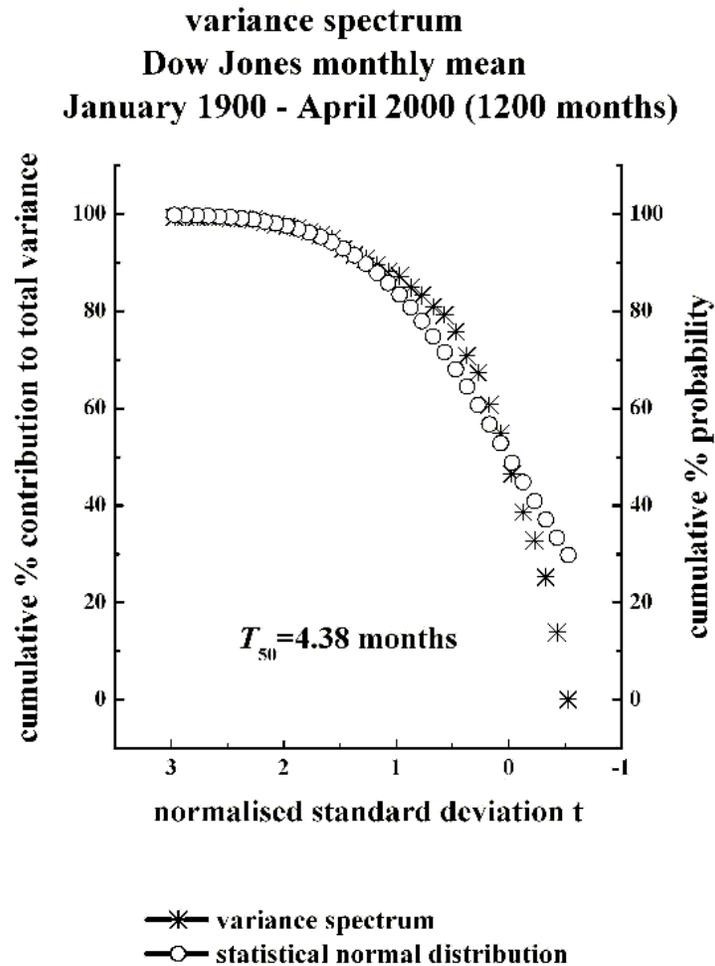

Figure 2

The number of dominant eddies with normalized variance $H \geq 1$ (see Section 4.2 above) is 199 out of which 24 are significant as determined by statistical $\chi^2$ test (see Section 4.1 above). The peak amplitude ($A_p$), the corresponding frequency ($f_p$), phase angle ($\phi_p$) and significance (sg) for each of the 199 dominant eddies are listed in Table 1.

Table 1: List of dominant eddies

| no | $L_p$ | $\phi_p$ | $A_p$ | $\nu_p$ | sg | no | $L_p$ | $\phi_p$ | $A_p$ | $\nu_p$ | sg |
|---|---|---|---|---|---|---|---|---|---|---|---|
| 1 | 2.008 | 66.786 | .027 | 3.586 |   | 55 | 2.757 | 75.466 | .015 | 1.740 |   |
| 2 | 2.038 | 196.026 | .024 | 2.356 |   | 56 | 2.799 | 69.852 | .017 | 1.716 |   |
| 3 | 2.047 | 225.634 | .020 | 4.689 |   | 57 | 2.807 | 334.909 | .015 | .855 |   |
| 4 | 2.057 | 219.532 | .014 | 1.167 |   | 58 | 2.821 | 319.754 | .014 | .851 |   |
| 5 | 2.061 | 132.148 | .019 | 1.165 |   | 59 | 2.855 | 288.038 | .014 | .841 |   |
| 6 | 2.092 | 30.095 | .016 | 2.293 |   | 60 | 2.881 | 124.679 | .014 | .833 |   |
| 7 | 2.115 | 242.991 | .023 | 4.541 |   | 61 | 2.898 | 167.626 | .015 | 1.657 |   |
| 8 | 2.126 | 218.663 | .027 | 4.509 |   | 62 | 2.913 | 264.820 | .021 | 2.472 |   |
| 9 | 2.141 | 275.771 | .016 | 1.121 |   | 63 | 2.927 | 150.484 | .015 | .820 |   |
| 10 | 2.147 | 121.855 | .020 | 2.237 |   | 64 | 2.954 | 14.780 | .016 | .812 |   |
| 11 | 2.158 | 108.341 | .023 | 4.451 |   | 65 | 2.963 | 249.840 | .017 | 1.621 |   |
| 12 | 2.167 | 183.411 | .025 | 4.429 | * | 66 | 2.975 | 101.123 | .026 | 2.420 |   |
| 13 | 2.175 | 127.031 | .014 | 1.103 |   | 67 | 2.999 | 33.906 | .016 | 1.600 |   |
| 14 | 2.180 | 156.057 | .014 | 2.201 |   | 68 | 3.014 | 248.143 | .014 | .796 |   |
| 15 | 2.186 | 78.559 | .018 | 1.098 |   | 69 | 3.026 | 92.111 | .016 | 1.586 |   |
| 16 | 2.199 | 52.319 | .031 | 2.181 | * | 70 | 3.087 | 189.980 | .022 | 2.332 |   |
| 17 | 2.215 | 306.603 | .033 | 5.424 | * | 71 | 3.099 | 118.122 | .018 | 1.548 |   |
| 18 | 2.226 | 201.453 | .024 | 3.238 |   | 72 | 3.149 | 28.291 | .018 | 1.525 |   |
| 19 | 2.235 | 109.007 | .014 | 1.074 |   | 73 | 3.162 | 268.009 | .018 | 2.277 |   |
| 20 | 2.248 | 53.530 | .022 | 1.068 |   | 74 | 3.174 | 138.421 | .014 | .756 |   |
| 21 | 2.266 | 17.475 | .026 | 2.119 |   | 75 | 3.200 | 200.794 | .018 | 2.250 |   |
| 22 | 2.287 | 187.173 | .017 | 2.098 |   | 76 | 3.216 | 9.246 | .014 | .746 |   |
| 23 | 2.314 | 74.611 | .016 | 1.037 |   | 77 | 3.324 | 19.284 | .014 | .722 |   |
| 24 | 2.319 | 316.439 | .016 | 2.069 |   | 78 | 3.341 | 249.589 | .015 | 1.438 |   |
| 25 | 2.366 | 336.401 | .033 | 4.060 | * | 79 | 3.401 | 74.847 | .016 | 1.411 |   |
| 26 | 2.373 | 185.218 | .024 | 1.011 |   | 80 | 3.425 | 353.454 | .016 | .701 |   |
| 27 | 2.380 | 84.271 | .015 | 2.018 |   | 81 | 3.470 | 336.278 | .022 | 2.765 |   |
| 28 | 2.390 | 131.550 | .016 | 1.004 |   | 82 | 3.508 | 234.698 | .022 | 2.052 |   |
| 29 | 2.411 | 8.248 | .019 | 1.992 |   | 83 | 3.522 | 127.010 | .018 | 1.362 |   |
| 30 | 2.421 | 141.863 | .023 | 6.926 |   | 84 | 3.540 | 286.577 | .016 | 1.355 |   |
| 31 | 2.448 | 166.168 | .028 | 2.942 | * | 85 | 3.558 | 64.977 | .017 | 1.350 |   |
| 32 | 2.455 | 33.032 | .024 | 1.954 |   | 86 | 3.590 | 124.070 | .017 | 1.336 |   |
| 33 | 2.462 | 308.061 | .014 | .975 |   | 87 | 3.622 | 115.706 | .022 | 5.308 |   |
| 34 | 2.480 | 183.081 | .021 | 1.935 |   | 88 | 3.666 | 16.652 | .021 | 3.276 |   |
| 35 | 2.502 | 30.937 | .014 | 1.917 |   | 89 | 3.710 | 300.441 | .019 | 1.940 |   |
| 36 | 2.520 | 338.323 | .032 | 2.857 | * | 90 | 3.770 | 60.803 | .029 | 3.183 | * |
| 37 | 2.530 | 125.271 | .021 | 2.846 |   | 91 | 3.797 | 122.606 | .018 | 1.896 |   |
| 38 | 2.540 | 9.423 | .016 | .945 |   | 92 | 3.889 | 282.058 | .014 | .617 |   |
| 39 | 2.548 | 313.691 | .018 | 1.885 |   | 93 | 3.948 | 5.481 | .032 | 4.256 | * |
| 40 | 2.555 | 167.315 | .020 | 1.878 |   | 94 | 3.999 | 319.638 | .029 | 2.399 | * |
| 41 | 2.568 | 4.724 | .015 | .935 |   | 95 | 4.023 | 252.452 | .014 | .596 |   |
| 42 | 2.576 | 342.624 | .024 | 2.792 |   | 96 | 4.052 | 231.569 | .021 | 1.777 |   |
| 43 | 2.596 | 349.082 | .019 | 1.850 |   | 97 | 4.121 | 302.914 | .014 | .582 |   |
| 44 | 2.617 | 265.272 | .019 | .917 |   | 98 | 4.142 | 201.781 | .017 | 1.738 |   |
| 45 | 2.625 | 196.914 | .021 | .914 |   | 99 | 4.183 | 225.163 | .028 | 4.012 | * |
| 46 | 2.633 | 99.837 | .014 | .911 |   | 100 | 4.230 | 180.653 | .016 | 1.135 |   |
| 47 | 2.665 | 241.385 | .016 | .901 |   | 101 | 4.259 | 256.316 | .026 | 2.255 |   |
| 48 | 2.676 | 46.712 | .014 | .897 |   | 102 | 4.285 | 150.979 | .027 | 3.359 | * |
| 49 | 2.684 | 320.596 | .015 | .894 |   | 103 | 4.320 | 245.703 | .021 | 2.221 |   |
| 50 | 2.692 | 252.161 | .016 | .892 |   | 104 | 4.350 | 331.040 | .022 | 2.761 |   |
| 51 | 2.702 | 162.455 | .026 | 4.436 |   | 105 | 4.376 | 170.736 | .030 | 2.742 | * |
| 52 | 2.719 | 36.033 | .014 | .883 |   | 106 | 4.438 | 65.898 | .018 | 3.247 |   |
| 53 | 2.732 | 282.403 | .016 | .878 |   | 107 | 4.460 | 319.841 | .016 | 1.614 |   |
| 54 | 2.749 | 194.703 | .021 | 2.619 |   | 108 | 4.487 | 173.438 | .016 | 1.605 |   |

| no | $L_p$ | $\phi_p$ | $A_p$ | $v_p$ | sg | no | $L_p$ | $\phi_p$ | $A_p$ | $v_p$ | sg |
|---|---|---|---|---|---|---|---|---|---|---|---|
| 109 | 4.523 | 196.545 | .025 | 2.124 |   | 156 | 9.283 | 251.581 | .025 | 2.327 |   |
| 110 | 4.555 | 42.169 | .025 | 3.689 | * | 157 | 9.470 | 2.140 | .022 | 2.791 |   |
| 111 | 4.596 | 131.764 | .023 | 5.727 |   | 158 | 9.623 | 93.107 | .018 | 1.746 |   |
| 112 | 4.651 | 136.604 | .023 | 2.580 |   | 159 | 9.857 | 90.877 | .015 | 3.671 |   |
| 113 | 4.750 | 140.476 | .017 | 4.052 |   | 160 | 10.116 | 327.361 | .015 | .949 |   |
| 114 | 4.783 | 358.806 | .014 | 1.004 |   | 161 | 10.228 | 276.626 | .018 | 1.173 |   |
| 115 | 4.851 | 275.142 | .018 | 1.484 |   | 162 | 10.624 | 187.389 | .018 | 1.356 |   |
| 116 | 4.880 | 175.803 | .016 | .983 |   | 163 | 11.203 | 76.876 | .025 | 3.004 |   |
| 117 | 4.994 | 197.495 | .018 | 1.442 |   | 164 | 11.372 | 19.426 | .015 | .845 |   |
| 118 | 5.024 | 153.195 | .019 | 1.433 |   | 165 | 11.955 | 62.281 | .020 | 1.607 |   |
| 119 | 5.079 | 59.632 | .032 | 2.837 | * | 166 | 12.124 | 344.747 | .030 | 2.178 | * |
| 120 | 5.125 | 174.592 | .019 | 3.278 |   | 167 | 12.543 | 239.154 | .016 | 1.530 |   |
| 121 | 5.260 | 27.807 | .018 | 1.369 |   | 168 | 12.873 | 319.355 | .019 | 1.678 |   |
| 122 | 5.318 | 318.186 | .017 | 1.354 |   | 169 | 13.226 | 286.472 | .022 | 2.359 |   |
| 123 | 5.372 | 39.263 | .023 | 2.682 |   | 170 | 14.313 | 336.497 | .026 | 2.180 |   |
| 124 | 5.442 | 4.846 | .024 | 3.973 |   | 171 | 14.529 | 258.379 | .022 | 1.817 |   |
| 125 | 5.602 | 319.885 | .031 | 3.863 | * | 172 | 14.942 | 274.720 | .016 | .964 |   |
| 126 | 5.641 | 216.856 | .020 | 1.701 |   | 173 | 15.929 | 154.168 | .025 | 2.110 |   |
| 127 | 5.692 | 330.655 | .020 | 1.686 |   | 174 | 16.267 | 61.284 | .028 | 2.506 | * |
| 128 | 5.727 | 211.835 | .022 | 2.513 |   | 175 | 16.612 | 328.178 | .015 | 1.444 |   |
| 129 | 5.837 | 191.903 | .021 | 2.466 |   | 176 | 17.135 | 266.753 | .027 | 3.218 | * |
| 130 | 5.901 | 143.397 | .017 | 1.220 |   | 177 | 18.378 | 318.920 | .019 | 1.959 |   |
| 131 | 5.943 | 42.565 | .040 | 5.634 | * | 178 | 19.262 | 246.129 | .019 | 2.116 |   |
| 132 | 6.111 | 111.207 | .024 | 2.752 |   | 179 | 21.267 | 226.889 | .020 | 1.693 |   |
| 133 | 6.160 | 352.700 | .015 | .779 |   | 180 | 22.627 | 147.812 | .021 | 3.308 |   |
| 134 | 6.198 | 255.139 | .021 | 1.936 |   | 181 | 24.002 | 143.018 | .031 | 2.500 | * |
| 135 | 6.310 | 200.088 | .019 | 4.175 |   | 182 | 25.872 | 60.367 | .024 | 2.225 |   |
| 136 | 6.393 | 299.845 | .018 | 1.877 |   | 183 | 28.223 | 17.831 | .023 | 3.327 |   |
| 137 | 6.470 | 319.357 | .015 | 1.113 |   | 184 | 29.759 | 151.839 | .021 | 2.738 |   |
| 138 | 6.528 | 179.501 | .021 | 2.940 |   | 185 | 31.949 | 89.588 | .016 | 1.127 |   |
| 139 | 6.774 | 154.906 | .019 | 1.416 |   | 186 | 33.554 | 290.346 | .023 | 5.254 |   |
| 140 | 7.009 | 138.538 | .014 | .342 |   | 187 | 36.935 | 307.947 | .021 | 1.950 |   |
| 141 | 7.150 | 49.218 | .014 | .671 |   | 188 | 40.131 | 216.488 | .036 | 7.147 | * |
| 142 | 7.229 | 214.466 | .027 | 2.655 |   | 189 | 46.625 | 114.593 | .023 | 1.852 |   |
| 143 | 7.412 | 136.100 | .024 | 4.837 |   | 190 | 49.657 | 63.623 | .020 | 1.978 |   |
| 144 | 7.570 | 88.245 | .020 | 6.306 |   | 191 | 53.205 | 100.375 | .018 | 1.715 |   |
| 145 | 7.754 | 91.401 | .033 | 3.405 | * | 192 | 57.234 | 25.366 | .017 | 1.753 |   |
| 146 | 7.839 | 237.026 | .014 | .918 |   | 193 | 63.889 | 85.148 | .022 | 7.458 |   |
| 147 | 8.094 | 18.918 | .030 | 5.041 | * | 194 | 91.941 | 199.328 | .017 | 2.017 |   |
| 148 | 8.249 | 323.318 | .016 | 1.163 |   | 195 | 119.839 | 199.585 | .025 | 5.047 | * |
| 149 | 8.341 | 187.344 | .020 | 1.727 |   | 196 | 145.496 | 3.594 | .014 | .478 |   |
| 150 | 8.425 | 96.630 | .019 | 1.708 |   | 197 | 179.316 | 201.317 | .017 | 1.269 |   |
| 151 | 8.492 | 101.529 | .015 | 1.131 |   | 198 | 415.776 | 227.996 | .019 | 3.439 |   |
| 152 | 8.569 | 41.615 | .019 | 1.400 |   | 199 | 2002.491 | 209.950 | .017 | 1.002 |   |
| 153 | 8.664 | 236.438 | .022 | 1.939 |   |   |   |   |   |   |   |
| 154 | 8.760 | 94.506 | .023 | 1.918 |   |   |   |   |   |   |   |
| 155 | 8.981 | 20.395 | .023 | 7.151 |   |   |   |   |   |   |   |

There are 24 significant peak periodicities marked as ∗ under column 'sg' in Table 1 and they correspond closely to model predicted dominant periodicities (Section 2.4, Equation 2). Spectral projection was done using all the 199 dominant eddies and also the 24 significant dominant eddies. The computed time series of 1266 normalized monthly mean Dow Jones Index values (January 1900 – October 2005) are shown in figure 3 along with the actual values for the years 2000 to 2005 and also for the historic years of major stock market crash during 1929 and 1987. The symbol D in figure 3 denotes the actual Dow Jones Index while the symbols A and S denote respectively spectral projections using all the dominant eddies and significant eddies. The times of occurrence of maxima and minima in Dow Jones Index values for the years 2000 to 2005 are in good agreement with the spectral projection.

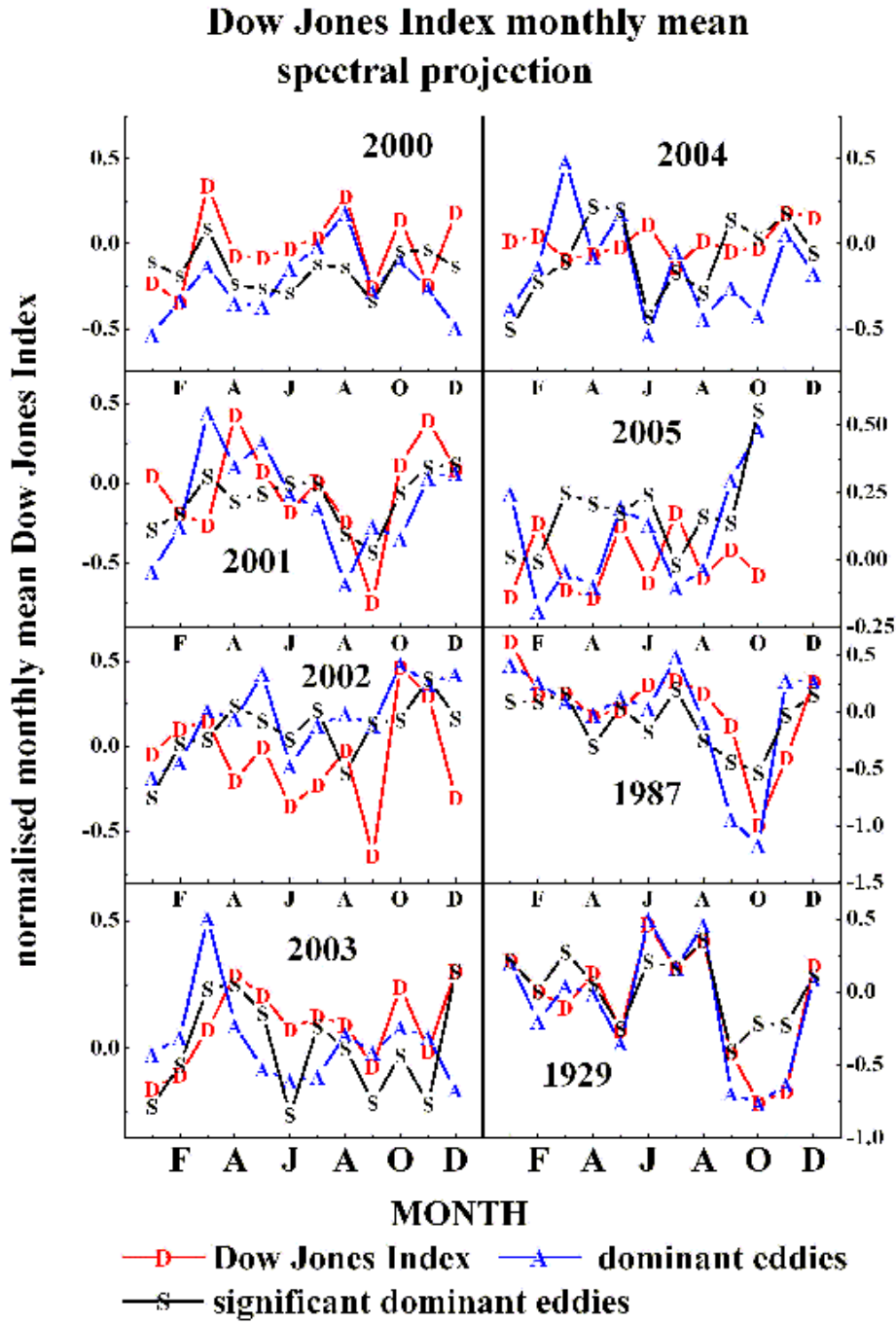

Figure 3

From a wide range of aggregate data, Chen (1996a) reports the existence of persistent cycles, in addition to color noise. Spectral analysis not only provides complementary evidence of "co-movements" of business, but also reveals distinctive patterns of frequency evolution. It is found that characteristic frequencies of business indicators are remarkably stable. Existence of stable characteristic frequencies from economic aggregates provides strong evidence of endogenous cycles and valuable information about structural changes. Economic behavior is more like an organism instead of random walks. Remarkable stability and resilience of market economy can be seen from the insignificance of the oil price shocks and the stock market crash. An alternative title for econometric literature could be: "Business cycle measurement without model specification". Chen (1996a) states that the most enlightening result in business cycle studies is the discovery of persistent cycles, i. e., self-generating cycles from economic aggregates. These cycles are nonlinear in nature with remarkable resilience and flexibility like living beings. This discovery provides a new perspective to business cycles. Traditionally, the economic order is characterized by negative feedback and equilibrium (steady) states. The new role of persistent cycles challenges the linear framework of economic dynamics. We need to re-examine the implications of complexity and instability in business cycles. The characteristic period of persistent cycles is around three to four years. The existence of persistent chaotic cycles reveals a new perspective of market resilience and new sources of economic uncertainties Chen (1996b).

The following spectral projection equation (see Section 4.2 and Equation 5) was used for the computation

$$\frac{1}{2}\sum_{p=1}^{npk} \nu_p A_p \cos(2\pi f_p \times time - \phi_p) = Y_{time}$$

where *npk* is the number of peak periodicities equal to 199 for all dominant periodicities and equal to 24 for significant periodicities. The computed data series $Y_{time}$ gives 1266 values for *time* = 1, 2, 3, …..1266. Spectral projection of near future fluctuations using dominant periodicities determined from historical (past) data series (1 to 1200 months in this study) is based on theory and model prediction (Equation 2) that dominant periodicities are functions of the golden mean and the time unit of measurement only and therefore remain the same for projection of the monthly mean data extending beyond 1200 months, e. g., up to 1266 months in the present study.

There is close agreement between historical and computed Dow Jones Index values for the first 1200 values which were used for the spectral analysis and thereafter there is reasonable agreement between the predicted and observed times of occurrence of maxima and minima (figure 3).

## Conclusion

The aggregate stock market activity as quantified in the Dow Jones Index exhibits fractal fluctuations on all time scales from days, months to years. Model predicted semi-permanent dominant periodicities were identified from continuous periodogram spectral analysis of available monthly mean historic data set from January 1900 - April 2000. These dominant periodicities were then used for spectral projection of fluctuation pattern for the period May 2000 to October 2005. The predicted and observed times of maxima and minima in Dow Jones Index were in close agreement (figure 3).

## Acknowledgement

The author is grateful to Dr. A. S. R. Murty for encouragement during the course of the study.